\documentclass[twocolumn,showpacs,prd,final]{revtex4}
\usepackage{amsmath}
\usepackage{graphicx}

\newcommand{\tr}{\mathop{\rm tr}}

\begin{document}

\title{Vacuum energy of two-dimensional $\mathcal{N}=(2,2)$ super
Yang-Mills theory}

\author{Issaku Kanamori}%
\email{kanamori@to.infn.it}
\affiliation{Theoretical Physics Laboratory, Nishina Center, Wako,
RIKEN, 351-0198, Japan\\
Dipartimento di Fisica Teorica, Universit\`a di Torino,
Via Giuria 1, 10125 Torino, Italy
}
\preprint{RIKEN-TH-149}

\begin{abstract}
We measure the vacuum energy of two-dimensional $\mathcal{N}=(2,2)$ 
super Yang-Mills theory using lattice simulation.
The obtained 
vacuum energy density is 
$\mathcal{E}_0=0.09(9)(^{+10}_{-8}) g^2$, where the first error is the
 systematic and the second is the statistical one, measured in
 the dimensionful gauge coupling~$g$ which governs the scale of the system.
The result is consistent with unbroken supersymmetry,
although we cannot exclude a possible very small non-zero vacuum energy.
\end{abstract}

\maketitle

\section{Introduction}

Lattice simulation of gauge theory is of great importance since it
provides a non-perturbative approach.
For the supersymmetric systems, 
both theoretical formulation on the lattice and simulation techniques 
are now rapidly developing (%
see Ref.~\cite{Giedt:2007hz} 
for a review and 
Refs.~\cite{
Catterall:2007fp,
Anagnostopoulos:2007fw,
Kanamori:2007ye,Kanamori:2007yx,
Elliott:2008jp,
Ishiki:2008te,
Giedt:2008xm,
Kikukawa:2008xw,
Catterall:2008dv,
Hanada:2008gy,
Kanamori:2008bk,
Kanamori:2008yy} 
for recent related works).
One of the mysterious points of supersymmetry (SUSY) 
is that it is broken in our universe, although it is widely believed that 
SUSY is a relevant symmetry beyond the standard model.
Because SUSY is such a strong symmetry, it is difficult to break it
spontaneously in perturbative calculations, 
unless it is broken at the tree level.  
The lattice simulation will help to understand 
the non-perturbative effects for the SUSY breaking.

In a recent work~\cite{Kanamori:2008bk}, 
on the basis of
a formulation by Sugino~\cite{Sugino:2004qd}
the author together with Suzuki explicitly confirmed that lattice artifacts
which break supersymmetry in fact disappear in the continuum
limit.
The target system was two-dimensional $\mathcal{N}=(2,2)$ super
Yang-Mills theory (for other lattice formulations of this system, 
see Refs.~\cite{
Kaplan:2002wv,Cohen:2003xe,Sugino:2003yb,Catterall:2004np,
Suzuki:2005dx,D'Adda:2005zk,Sugino:2006uf}).
In order to suppress the fluctuation along the flat direction, we added
scalar mass terms and numerically observed the Partially Conserved 
SuperCurrent (PCSC) relation, 
which is the conservation law of the supercurrent modified by
the scalar mass.  
Furthermore, we illustrated some physical applications,
power-like behavior of certain correlation functions and static
potential between fundamental probe charges~\cite{Kanamori:2008yy}.

The vacuum energy is the order parameter for SUSY
breaking~\cite{Witten:1981nf}
and we are interested in the spontaneous SUSY breaking of this
system.  Hori and Tong~\cite{Hori:2006dk} pointed out the
possibility of spontaneous breaking
while the previous simulation presented 
in Refs.~\cite{Kanamori:2007ye,Kanamori:2007yx} was consistent with
no SUSY  breaking.
The vacuum energy obtained by the above simulation, 
however, had a rather large error because of
an inefficient simulation algorithm.

The purpose of this letter is to measure the vacuum energy with a small
error which would help to determine whether supersymmetry is
spontaneously broken or not in this system.
We need, of course, to separate the spontaneous supersymmetry breaking 
from a breaking caused by lattice artifacts.
The result in Ref.~\cite{Kanamori:2008bk} shows that
the latter does not survive in the continuum limit.

The target two-dimensional system is a nice arena for simulation of
supersymmetric systems,
although it is a toy system from a viewpoint of phenomenologically 
realistic four dimensional models.  
Since it is two-dimensional, computational cost is less than
four-dimensional case.  Furthermore, 
there is no sign problem in the continuum target theory, which
also makes simulation easy.
Important differences from four-dimensional $\mathcal{N}=1$
supersymmetric Yang-Mills~\footnote{
See, for example, Ref.~\cite{Giedt:2009yd} 
for a recent report and references there in.
} are the existence of massless scalar fields and
the existence of flat directions in the classical potential.

\section{Formulation}

We use the lattice model in Ref.~\cite{Sugino:2004qd}
and the method proposed 
in Refs~\cite{Kanamori:2007ye, Kanamori:2007yx} to measure the vacuum energy.

The action of the target $\mathcal{N}=(2,2)$ super Yang-Mills theory
is obtained by dimensional reduction from 4-dimensional
$\mathcal{N}=1$ super Yang-Mills theory.
In the Euclidean continuum spacetime, it is
\begin{align}
  S&=\frac{1}{g^2}
   \int d^2x\,\tr\left\{
   \frac{1}{2}F_{MN}F_{MN}+\Psi^TC\Gamma_MD_M\Psi+\widetilde H^2\right\}.
 \label{eq:continuum_action}
\end{align}
Here, $M$ and $N$ run from $0$ to $3$. $\Gamma_M$ are the 
gamma matrices in four dimensions and $C$ is the charge conjugation matrix.
The covariant derivative $D_M$ is $\partial_M+i[A_M, \cdot]$, where
$\partial_2=\partial_3=0$, $A_0$ and $A_1$ are two-dimensional gauge
fields, $A_2$ and $A_3$ are real scalar fields.
The field strength is $F_{MN}=\partial_M A_N-\partial_N A_M+i[A_M,A_N]$.
The fermion $\Psi=(\psi_0, \psi_1, \chi, \eta/2)^T$ is a 4 component spinor
and $\widetilde{H}$ is a bosonic auxiliary field.
The gauge coupling $g$ has mass dimension $1$ and we use it
as a unit for dimensionful quantities.
The action is invariant under four 
super transformations 
and one can rewrite it
in a supercharge exact form.

The lattice action we use
is $ S_{\rm lattice}=S_{\rm Sugino} + S_{\rm scalar\ mass}$~\cite{Kanamori:2008bk},
where $S_{\rm Sugino}$ is the lattice version of 
(\ref{eq:continuum_action}) and has one exact nilpotent 
symmetry $Q$ at finite lattice spacings out of four 
supercharges~\cite{Sugino:2004qd}. 
$S_{\rm scalar\ mass}$, which softly breaks the supersymmetry as well as
$Q$-symmetry,
is needed to lift the flat direction of the
potential, because otherwise the simulation does not give thermalized
configurations \cite{Kanamori:2008vi}.  
We thus \emph{define} the system as an extrapolation in which the
scalar mass~$\mu$ goes to~$0$ after taking continuum limit.
In taking the continuum limit, we use fixed scalar mass in a physical unit.

The vacuum energy is the order parameter for SUSY breaking.
It is positive
if and only if SUSY is spontaneously broken.
Therefore it is of crucial importance 
to choose the correct origin of the energy.  
Our choice is to
use a $Q$-exact Hamiltonian (density). 
In Refs.~\cite{Kanamori:2007ye, Kanamori:2007yx}, 
the author together with his collaborators
discussed that this choice is consistent with a property of the
Witten index.
They also confirmed that this
method reproduces known results for supersymmetric quantum mechanics.
For the current system, the Hamiltonian density is
 $\mathcal{H}=\frac{1}{2}Q\mathcal{J}_0^{(0)} $,
where $\mathcal{J}_0^{(0)}$ is the 0-th component of the Noether current 
corresponding to another supercharge $Q_0$.  
Note that $Q$ and $Q_0$ satisfy the super algebra
$\{Q,Q_0\}=-2i\partial_0$ in the continuum.
Since we do not have a
lattice version of the $Q_0$~transformation, we use a naive discretization
of the continuum current (see Ref.~\cite{Kanamori:2007yx} for its explicit
expression).
We do not include the scalar mass term
in the Hamiltonian density because the Noether current
$\mathcal{J}^{(0)}$ does not contain it.
The Hamiltonian density is exactly identical to the one treated in
Refs~\cite{Kanamori:2007ye, Kanamori:2007yx}.

We 
utilize the temperature as a conjugate external field to the Hamiltonian,
i.e., we impose antiperiodic boundary condition (aPBC) 
for fermion fields in time direction.  
We use a finite spacial size $L_{\rm S}$ and temporal size $\beta$,
where $\beta$ is the inverse temperature as well.
What we measure is
\begin{equation}
 \langle\mathcal{H}\rangle
 =\frac{\int_{\rm aPBC}{\rm d} \mu\, \mathcal{H} \exp(-S_{\rm lattice})
 }{\int_{\rm aPBC}{\rm d} \mu\, \exp(-S_{\rm lattice})},
 \label{eq:expectation-value}
\end{equation}
where we explicitly indicate the boundary condition.
${\rm d} \mu$ denotes 
an integration measure of the path integral.
If we adopted the periodic boundary condition, since 
the expectation value of Hamiltonian would be proportional to a derivative
of the Witten index w.r.t the coupling constant, the expectation value
would be always
zero~\footnote{
In the case of 
periodic boundary condition,
if SUSY is broken, the partition function becomes zero and the
expectation value will be $\frac{0}{0}$.
Numerically, one will
encounter a severe sign problem caused by sign (or complex phase) of the
Pfaffian.}. 
Therefore, we could not obtain useful information for the SUSY breaking.
We will take the zero temperature limit at the final stage to obtain the
vacuum energy density:
$ \mathcal{E}_0
  \equiv\lim_{\beta\to\infty}(\lim_{\mu\to 0}(\lim_{a\to 0} \langle
  \mathcal{H}\rangle))$~\footnote{
In the extrapolation $\mu\to 0$, we use the values at 
$\mu=1.3g,\ 1.0g,\ 0.7g$ and $0.5g$ for all $\beta$.  
The Hamiltonian density is smoothly extrapolated
for each $\beta$. See Fig.~\ref{fig:massless} and related descriptions.
}.

\section{Simulation}
We evaluate the expectation value~(\ref{eq:expectation-value}) 
using a Monte Carlo simulation.
The detail of generating the configurations is found 
in Refs.~\cite{Kanamori:2008bk,Kanamori:2008vi}
\footnote{
The code was developed using FermiQCD/MDP~\cite{mdp}
and the parameters for rational expansion were obtained from a program
in Ref.~\cite{Remez}.
}.
We use the Rational Hybrid Monte Carlo (RHMC) method 
together with the multi-time step 
for molecular dynamics.
Compared with the method used in~\cite{Kanamori:2007ye,
Kanamori:2007yx}, which uses quenched configurations and 
reweighting method to treat the fermion effect, 
the statistical error is well under control.

\begin{figure}
 \includegraphics[
 width=0.9\linewidth
,height=0.55\linewidth
 ]{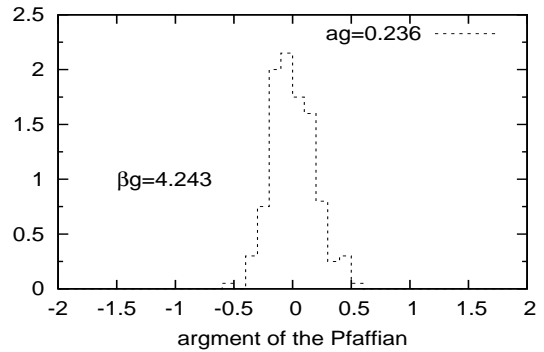}
\caption{Distribution of the argument of the Pfaffian for 200 samples.
The scalar mass is $\mu=0.5g$.
The argument is centered around $0$, which justifies the replacement of
 the Pfaffian by the positive square root of the determinant.
}
\label{fig:pfs}
 \end{figure}

\begin{figure}
 \hfil
 \includegraphics[
 width=0.9\linewidth
,height=0.55\linewidth
 ]{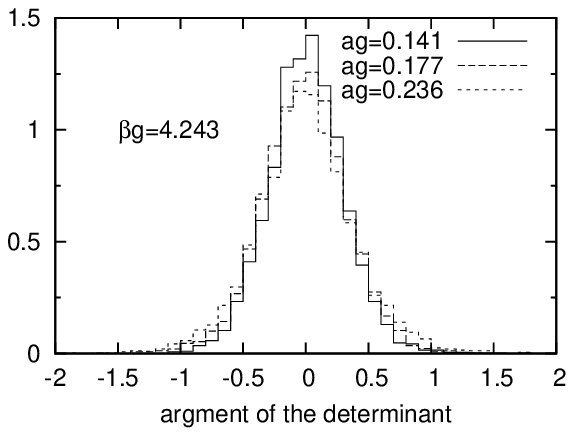}\\
 \hfil
 \includegraphics[
 width=0.9\linewidth
,height=0.55\linewidth
 ]{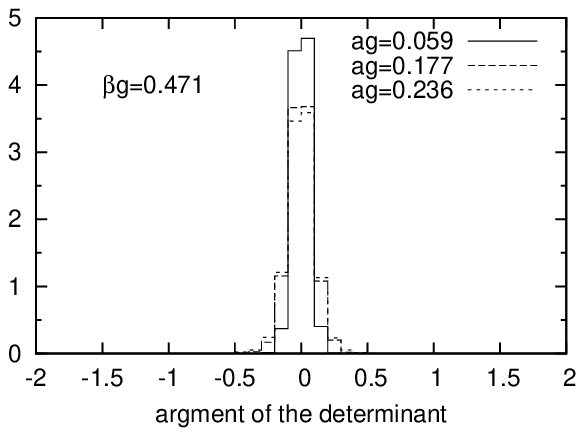}
 \hfil
 \caption{Distributions of the argument of the determinant.  The upper
 plot is one of the broadest distribution and the lower is one of the
 narrowest.  For each plot, the scalar mass is $\mu=0.5g$.  
 Plot with $\beta g=4.243$, $ag=0.236$ uses the same parameter 
 as the plot in Fig.~\ref{fig:pfs}.
 Even in the
 broadest cases, the argument is centered around zero.}
 \label{fig:dets}
\end{figure}

The effect of the complex phase of the Pfaffian of the Dirac operator
is taken into account by phase reweighting of the
square root of the determinant.
Note that the target system in the continuum has real positive Pfaffian.
Because our choice of the lattice spacing is small enough and the
argument of the determinants is distributed around zero,
there is almost no sign ambiguities caused by taking the square root.
In Fig.~\ref{fig:pfs} and Fig.~\ref{fig:dets}, 
we plot the distribution of the argument 
of the determinant and a sample distribution from a direction
calculation of the Pfaffian~\footnote{
The author thanks to H.~Suzuki, for letting the author use his code
for calculating the Pfaffian.
}.
Fig.~\ref{fig:pfs} shows
that the Pfaffian is almost real and positive~\footnote{
Although this results is quite reasonable since it is consistent with 
the real positiveness of the Pfaffian of the continuum theory,
it is rather different from the result in
Ref.~\cite{Catterall:2008dv}.
Important differences from Ref.~\cite{Catterall:2008dv} 
are lattice model used in the
simulation, boundary condition and system size.  One (or more) of them
may explain the difference.
} even for the
largest lattice spacing and the lowest temperature, and guarantees the
replacement of Pfaffian by the positive (i.e., the real part
is positive) square root of the determinant.
Note that the direct calculation of the Pfaffian is numerically very
expensive.
From Fig.~\ref{fig:dets}, one can see that as the lattice spacing
becomes small the determinant and hence the Pfaffian approaches real and
positive values, which is consistent with the property in the continuum.

In the measurement, we calculate the Hamiltonian density every~10
trajectories.  We use binning technique (a typical bin size is~5)
and use the jack-knife method to estimate the statistical error for each
parameters.  The continuum limit is taken as a linear extrapolation
w.r.t. the lattice spacing~$a$.  The massless extrapolation is taken as a linear
extrapolation w.r.t.~$\mu^2$.  
We use 
$\chi^2$ fit and estimate the statistical error of the fitting parameters
as the value for which 
$\chi^2$ increases by $1$.

The common parameters are the following.
We use SU(2) gauge group.  Physical spacial size~$L_{\rm S}$ is fixed to
$L_{\rm S}g=1.414$.
The parameter $\epsilon$ for the admissibility condition in $S_{\rm Sugino}$
 is $2.6$.

\section{Results}

We list the detailed parameters and raw results in Table~\ref{table:table}.
For large~$\beta$, we use 3 different lattice spacings to obtain the
continuum limit while for small~$\beta$ we use 4 or more.
It is because that in the former cases the obtained values are
almost constant within the statistical error over the different
lattice spacings; 
in the latter cases they depend on the lattice spacing.

\begin{figure}
 \hfil
 \includegraphics[width=0.9\linewidth
,height=0.55\linewidth
]{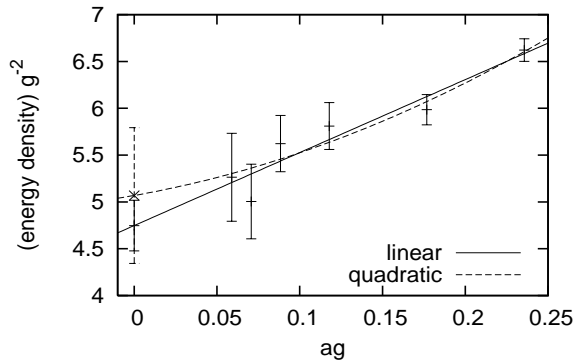}
\caption{Example of the continuum limit, for $\beta g=0.707$ 
and~$\mu=0.7g$.
 We use data set with $\beta g=0.707$ to estimate systematic errors
 associated with the choice of the fitting function and the fitting region.
 }
\label{fig:cont_limit}
\end{figure}

We estimate a systematic error in taking the continuum limit using
the data set with $\beta=0.707g$ (Fig.~\ref{fig:cont_limit}).  
Although we adopt the linear fitting 
to obtain the continuum limit,
we tried 
a quadratic function $A'a^2+B'a+C'$ as well as a linear one
$A a +B$ in the lattice spacing~$a$.
After extrapolating $\mu\to 0$,
we obtained two different values and we regard the difference as the
systematic error in choosing the fitting function, which is 4.0\%.
To estimate the errors associated with a choice of the fitting region, 
we repeated a similar analysis with linear fitting
using small 4 lattice spacings and
large 3 lattice spacings.  Comparing these two results with the one
obtained from all 6 lattice spacings, we estimate the error associated with
the choice of the fitting region at 6.8\%.
In total, the systematic error in taking the continuum
limit is 11\%.  

\begin{figure}
 \hfil
 \includegraphics[width=0.9\linewidth
 ,height=0.55\linewidth
 ]{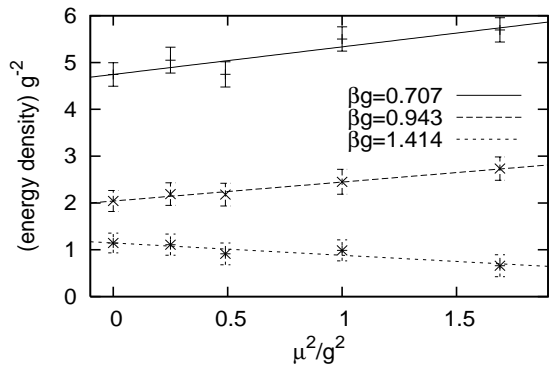}
 \caption{Examples of the extrapolation of the scalar mass $\mu$ to~0.
  The linear extrapolations in $\mu^2$ fit very well.}
 \label{fig:massless}
\end{figure}

After taking the continuum limit, we extrapolate the scalar 
mass~$\mu\to 0$
(Fig.~\ref{fig:massless}).  
As the figure indicates, a possible systematic error in the
extrapolation is small compared with the statistical ones 
and thus is negligible.

\begin{figure}
  \hfil
 \includegraphics[width=0.9\linewidth
 ,height=0.55\linewidth
 ]{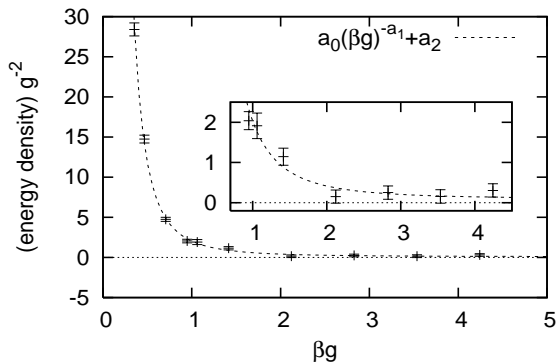}
 \caption{Energy density $\mathcal{E}(\beta)$. 
 Errors bars denote statistical errors only.
 The dashed curve is the best fit of data with a power function of $\beta$.
 As the inverse
 temperature $\beta$ becomes large, the energy density approaches $0$.}
 \label{fig:energy}
\end{figure}

The final result is given in Fig.~\ref{fig:energy}.
We have fitted the result using two different functions: a power
of~$\beta$, $\mathcal{E}(\beta)/g^2= a_0 (\beta g)^{-a_1}+a_2$,
and an exponential of~$\beta$, 
$\mathcal{E}(\beta)/g^2= b_0 \exp(-b_1 \beta g) +b_2$. 
We obtain
  $a_0=1.94^{{}+11}_{{}-16}$,
  $a_1=2.59^{{}+9}_{{}-6}$, 
  $a_2=0.09^{{}+10}_{{}-8}$ 
and 
  $b_0=157^{{}+13}_{{}-12}$, 
  $b_1=5.01^{{}+18}_{{}-17}$, 
  $b_2=0.37^{{}+8}_{{}-8}$.  
The errors are only statistical.  
The values of $\chi^2$ per degrees of freedom are $1.21$ and
$3.81$ for 7 degrees of freedom, respectively.
Using the result of the fit by power function,
we obtain $\mathcal{E}_0=0.09^{{}+10}_{{}-8} g^2$ for the central value 
and the statistical errors. 

However, because the above fits use small~$\beta$ as well as large~$\beta$, 
one cannot exclude possible
effects from the excited states.  In fact, using only large 
$\beta$ ($\beta g>0.9$), we obtain reliable fits for both exponential and
power, which gives 
  $\mathcal{E}_0=0.17^{{}+10}_{{}-11}g^2$ 
and 
  $\mathcal{E}_0=0.05^{{}+17}_{{}-17}g^2$,
respectively~%
\footnote{
Other coefficients are $a_0=1.9^{{}+2}_{{}-2},\ a_1=2.2^{+{}9}_{{}-5}$
and $b_0=13^{{}+9}_{{}-5},\ b_1=2.0^{{}+6}_{{}-4}$.
}.  
These values are consistent with the above central value within the
statistical error. 
Using these values,
we estimate that the systematic error in taking zero temperature limit
is $0.08 g^2$.

What is the physical difference between the power and the exponential fitting?
If the spectrum has a gap, 
the energy behaves as an exponential function of $\beta$.
A power behavior of $\beta$ comes from gap-less excitations
 \footnote{
If the energy density function behaves as 
$\rho(\mathcal{E})\sim \mathcal{E}^{\nu-1}$,
$\mathcal{E}(\beta)$ behaves as $\nu/\beta$ for large $\beta$
which gives $a_1=1.0$.
}.
In a finite volume system, usually we have 
a discrete spectrum which leads to a mass gap.
However, the current system has the \emph{flat direction} in the potential
at least in the classical level,
and the fluctuation along the flat direction can provide 
a continuum spectrum.
The current result does not allow us to determine which is the
case~\footnote{
Although 
$a_1=2.2^{+{}9}_{{}-5}$
for $\beta g>0.9$ is rather different from~$1.0$ and seems to 
prefer the exponential fit,
we cannot totally exclude the possibility of power function with the
current error.%
}.
If the fit by the power was 
preferred, 
it would be an evidence of
the recovery of the flat direction after the $\mu\to 0$ 
extrapolation.

In total, we obtain the vacuum energy
  $\mathcal{E}_0=0.09(9)(^{{}+10}_{{}-8}) g^2$,
where the first error is the systematic and the second is the
statistical one.
It is consistent with $0$ within the error, that is, 
it is consistent with
supersymmetry not being spontaneously broken.

\begin{table*}
\caption{
 Expectation value of the Hamiltonian density
 $\langle\mathcal{H}\rangle$ and number of the independent
 configurations after binning for each parameters. 
 $g$ is the dimensionful gauge coupling.
}
\label{table:table}
\begin{tabular}{c@{\hspace{1em}}cc@{\hspace{1.7em}}cc@{\hspace{1.7em}}cc@{\hspace{1.7em}}cc@{\hspace{1.7em}}cc}
\hline\hline
&\multicolumn{2}{c}{}&\multicolumn{2}{c}{$ \mu=0.5g $}&\multicolumn{2}{c}{$ \mu=0.7g $}&\multicolumn{2}{c}{$ \mu=1.0g $}&\multicolumn{2}{c}{$ \mu=1.3g $}\\
 $\beta g$ & $ag$ & lat.~size  
	 & num. & $\langle \mathcal{H} \rangle g^{-2}$ 
	 & num. & $\langle \mathcal{H} \rangle g^{-2}$ 
         & num. & $\langle \mathcal{H} \rangle g^{-2}$ 
         & num. & $\langle \mathcal{H} \rangle g^{-2}$ \\
\hline
0.354 & 0.118 & $ 3 \times 12$ & 800 & 34.50(32) & 800 & 34.39(32) & 2000 & 33.57(31) & 800 & 33.55(34) \\
 & 0.088 & $ 4 \times 16$ & 1200 & 32.66(35) & 1200 & 32.45(35) & 3000 & 32.49(34) & 3000 & 31.59(35) \\
 & 0.071 & $ 5 \times 20$ & 1600 & 31.98(39) & 1600 & 31.99(38) & 2000 & 30.89(38) & 800 & 30.60(40) \\
 & 0.059 & $ 6 \times 24$ & 400 & 31.47(66) & 800 & 30.54(65) & 400 & 29.80(67) & 800 & 29.72(65) \\
\hline
0.471 & 0.157 & $ 3 \times 9$ & 800 & 17.59(24) & 800 & 17.89(22) & 800 & 17.04(21) & 1000 & 17.62(21) \\
 & 0.118 & $ 4 \times 12$ & 1000 & 17.06(28) & 400 & 16.66(30) & 800 & 16.58(28) & 1000 & 16.28(28) \\
 & 0.079 & $ 6 \times 18$ & 800 & 16.03(32) & 1600 & 16.15(30) & 1600 & 16.31(30) & 1600 & 15.83(30) \\
 & 0.059 & $ 8 \times 24$ & 2000 & 16.05(53) & 2000 & 15.64(56) & 2000 & 15.75(56) & 500 & 16.49(58) \\
\hline
0.707 & 0.236 & $ 3 \times 6$ & 400 & 6.18(12) & 800 & 6.62(12) & 800 & 6.64(11) & 800 & 6.51(11) \\
 & 0.177 & $ 4 \times 8$ & 200 & 5.77(17) & 800 & 5.99(16) & 800 & 6.40(15) & 800 & 6.18(16) \\
 & 0.118 & $ 6 \times 12$ & 800 & 5.70(23) & 500 & 5.81(24) & 2000 & 5.87(22) & 400 & 5.85(23) \\
 & 0.088 & $ 8 \times 16$ & 400 & 5.43(32) & 800 & 5.62(30) & 800 & 5.89(30) & 400 & 6.15(29) \\
 & 0.071 & $ 10 \times 20$ & 400 & 4.87(41) & 800 & 5.00(39) & 400 & 5.87(38) & 400 & 5.94(37) \\
 & 0.059 & $ 12 \times 24$ & 800 & 5.95(45) & 400 & 5.26(46) & 400 & 6.27(46) & 2000 & 6.36(45) \\
\hline
0.943 & 0.157 & $ 6 \times 9$ & 800 & 2.62(15) & 800 & 2.65(15) & 800 & 2.93(15) & 800 & 2.96(15) \\
 & 0.118 & $ 8 \times 12$ & 800 & 2.54(19) & 800 & 2.73(19) & 800 & 2.75(19) & 800 & 2.61(20) \\
 & 0.079 & $ 12 \times 18$ & 1600 & 2.07(21) & 1600 & 2.35(22) & 1600 & 2.62(21) & 800 & 2.72(22) \\
\hline
1.061 & 0.177 & $ 6 \times 8$ & 800 & 1.70(12) & 400 & 1.97(13) & 800 & 2.09(12) & 400 & 1.97(12) \\
 & 0.118 & $ 9 \times 12$ & 2000 & 1.74(17) & 800 & 2.04(18) & 2000 & 2.05(18) & 800 & 2.46(17) \\
 & 0.088 & $ 12 \times 16$ & 800 & 1.94(18) & 1000 & 1.85(18) & 800 & 1.82(18) & 1600 & 1.75(17) \\
\hline
1.414 & 0.236 & $ 6 \times 6$ & 800 & 0.777(76) & 400 & 0.918(81) & 800 & 0.961(77) & 2000 & 1.046(75) \\
 & 0.177 & $ 8 \times 8$ & 800 & 1.00(10) & 400 & 1.07(10) & 1000 & 1.08(10) & 800 & 0.97(10) \\
 & 0.118 & $ 12 \times 12$ & 1600 & 0.89(11) & 800 & 0.86(11) & 1600 & 0.93(11) & 800 & 0.84(11) \\
\hline
2.121 & 0.236 & $ 9 \times 6$ & 800 & 0.471(66) & 2000 & 0.542(63) & 2000 & 0.454(61) & 2000 & 0.435(60) \\
 & 0.177 & $ 12 \times 8$ & 2000 & 0.377(82) & 800 & 0.574(88) & 800 & 0.491(86) & 2000 & 0.548(82) \\
 & 0.118 & $ 18 \times 12$ & 800 & 0.442(93) & 1333 & 0.319(94) & 1500 & 0.55(10) & 1200 & 0.52(10) \\
\hline
2.828 & 0.236 & $ 12 \times 6$ & 800 & 0.349(57) & 2000 & 0.417(53) & 2000 & 0.442(54) & 4000 & 0.385(52) \\
 & 0.177 & $ 16 \times 8$ & 800 & 0.339(78) & 800 & 0.477(75) & 2000 & 0.280(74) & 800 & 0.471(75) \\
 & 0.118 & $ 24 \times 12$ & 1200 & 0.349(94) & 1500 & 0.362(90) & 300 & 0.22(10) & 1500 & 0.412(94) \\
\hline
3.536 & 0.236 & $ 15 \times 6$ & 800 & 0.410(49) & 800 & 0.246(48) & 1000 & 0.326(49) & 800 & 0.370(49) \\
 & 0.177 & $ 20 \times 8$ & 400 & 0.359(68) & 500 & 0.366(68) & 800 & 0.346(72) & 400 & 0.379(72) \\
 & 0.118 & $ 30 \times 12$ & 2000 & 0.28(10) & 1000 & 0.20(10) & 1000 & 0.300(97) & 1000 & 0.439(98) \\
\hline
4.243 & 0.236 & $ 18 \times 6$ & 2000 & 0.286(44) & 800 & 0.255(44) & 4000 & 0.302(42) & 400 & 0.346(43) \\
 & 0.177 & $ 24 \times 8$ & 800 & 0.290(62) & 400 & 0.307(68) & 2000 & 0.310(60) & 800 & 0.292(64) \\
 & 0.141 & $ 30 \times 10$ & 800 & 0.241(80) & 800 & 0.278(78) & 400 & 0.382(79) & 800 & 0.339(78) \\
\hline\hline
\end{tabular}
\end{table*}

\section{Conclusion}
In this letter, we measured the vacuum energy of
two-dimensional $\mathcal{N}=(2,2)$ super Yang-Mills theory
using lattice simulation.
Because our simulation does not have explicit supersymmetry breakings
 caused by lattice artifacts, the vacuum energy we measured 
works as an order parameter of \emph{spontaneous} SUSY breaking.
The result is consistent with~$0$
within the error, 
which seems to imply that
supersymmetry is not spontaneously broken,
although we cannot exclude the possibility of a non-zero value
smaller than the error which would indicate a spontaneous breaking of SUSY.

\acknowledgments
I.~K. is supported by the Special Postdoctoral Research Program at
RIKEN.  He is also supported by the Nishina Memorial Foundation.
He thanks H.~Suzuki for encouragement and discussion.
He also thanks A.~D'Adda, A.~Fotopoulos, M.~Hanada, D.~Kadoh, 
N.~Kawamoto, Y.~Kikukawa and F.~Sugino for various comments 
and discussion.
The simulation was performed on the RIKEN Super
Combined Cluster.

\end{document}